# DID THE TERRESTRIAL PLANETS OF THE SOLAR SYSTEM FORM BY PEBBLE ACCRETION?

A. Morbidelli[1,2], T. Kleine[3] and F. Nimmo[4]

[1.] Collège de France, CNRS, PSL Univ., Sorbonne Univ., Paris, 75014, France

[2.] Laboratoire Lagrange, Université Cote d'Azur, CNRS, Observatoire de la Côte d'Azur, Boulevard de l'Observatoire, 06304 Nice Cedex 4, France

[3.] Max Planck Institute for Solar System Research, Justus-von-Liebig-Weg 3, 37077 Göttingen, Germany

[4.] Dept. Earth and Planetary Sciences, University of California Santa Cruz, Santa Cruz CA 95060, USA

**Abstract**

The dominant accretion process leading to the formation of the terrestrial planets of the Solar System is a subject of intense scientific debate. Two radically different scenarios have been proposed. The classic scenario starts from a disk of planetesimals which, by mutual collisions, produce a set of Moon to Mars-mass planetary embryos. After the removal of gas from the disk, the embryos experience mutual giant impacts which, together with the accretion of additional planetesimals, lead to the formation of the terrestrial planets on a timescale of tens of millions of years. In the alternative, pebble accretion scenario, the terrestrial planets grow by accreting sunward-drifting mm-cm sized particles from the outer disk. The planets all form within the lifetime of the disk, with the sole exception of Earth, which undergoes a single post-disk giant impact with Theia (a fifth protoplanet formed by pebble accretion itself) to form the Moon. To distinguish between these two scenarios, we revisit all available constraints: compositional (in terms of nucleosynthetic isotope anomalies and chemical composition), dynamical and chronological. We find that the pebble accretion scenario is unable to match these constraints in a self-consistent manner, unlike the classic scenario.

## 1. Introduction

In the classic model, planet formation starts from a disk of planetesimals, i.e., km-sized objects akin to asteroids or Kuiper belt objects. Through phases of runaway and oligarchic growth by mutual planetesimal collisions (Greenberg et al., 1978; Kokubo and Ida, 1998), a few dozens of large objects form, called planetary embryos, resulting in a strongly bimodal size distribution. After the removal of gas from the disk, the embryos continue to grow by mutual giant impacts and accretion of planetesimals, forming bodies of ever-increasing size, ultimately resulting in the formation of the terrestrial planets on timescales of tens of millions of years (My). This process is quite successful in reproducing the main properties of the terrestrial planets of our Solar System (e.g. Raymond et al., 2006), particularly if the initial population of planetesimals is assumed to be confined in a narrow range of radial distances (Hansen, 2009; Nesvorny et al., 2021;

Woo et al., 2023). However, the classical model fails dramatically to form the cores of the giant planets, namely objects massive enough to be able to capture gas from the protoplanetary disk before its dispersal (i.e. within a few My; Levison et al., 2010)

Pebble accretion offers a brilliant solution to this problem of the rapid formation of the giant planets' cores. Pebble accretion is a growth process by which a protoplanet accretes macroscopic dust particles, dubbed pebbles, as they drift towards the star in the protoplanetary disk due to gas drag (Ormel and Klahr, 2010; Lambrechts and Johansen, 2012; Ida et al., 2016 and many others). The particles that are effective in this process have Stokes numbers (i.e., the product between the orbital frequency and stopping time due to gas drag) typically between 0.001 and 0.1 (which corresponds to mm-cm sizes depending on the local gas density of the disk). We will use dust and pebbles as synonymous in this manuscript. Due to a combination of gravitational deflection and gas drag, the planet's accretional cross section for these particles is significantly larger than that for planetesimals, which are large enough to be immune to gas drag during a close encounter with the planet. Moreover, while a planet can only be fed from a local population of planetesimals, dust particles are radially mobile and thus the planet can in principle grow from a much larger reservoir of material. Under reasonable conditions, it has been shown that the cores of the giant planets of the Solar System could have formed within the disk's lifetime of a few My by the pebble accretion process (see Drazkowska et al., 2023 for a review). Yet, the discovery of dust rings in protoplanetary disks by ALMA (Andrews et al., 2018) casts some doubt that dust can freely move through the disk as initially envisioned in the pebble accretion model, but it has been shown that a giant planet's core can still rapidly accrete from a ring of dust (Velasco-Romero et al., 2024).

Given the success of the pebble accretion model in explaining the formation of giant planets, it is natural to propose that pebble accretion is the dominant process leading to the formation of any planet, including the terrestrial planets of the Solar System (Johansen et al., 2021). The efficiency of pebble accretion, however, depends on two main parameters: (*i*) the availability of a sufficiently large reservoir of pebbles and (*ii*) their concentration near the mid-plane, where the sedimentation of pebbles depends on the particles' Stokes number and the turbulent diffusion in the disk. Pebble accretion can therefore be efficient or inefficient depending on these parameters. This makes the structure of the gas-dust disk the key parameter determining whether a given set of planets grows predominantly from a drifting mass of pebbles or from planetesimal impacts and merging events with other protoplanets (a.k.a. giant impacts).

The initial structure of the solar accretion disk is not known a priori, and so assessing whether the terrestrial planets of the Solar System formed by pebble accretion or by the classic model is only possible by comparing model predictions to observational constraints. This is precisely the aim of this work: we consider the widest possible set of constraints (compositional, dynamical, chronological) to assess whether pebble accretion is a possible process for the growth of the rocky planets of the Solar System or whether it can be ruled out. Some of our considerations have already been presented in previous works, but always in a scattered form and often being treated in a piecemeal fashion within the pebble accretion paradigm. However, we believe it is important to follow a holistic approach, in which all constraints are considered together in a self-consistent manner to ultimately assess whether they can be comprehensively satisfied by either the pebble accretion process or the classic collisional growth model, or a combination thereof. To this end it is important to emphasize that nothing is wrong with the pebble accretion process *per se*; the question rather is whether the protosolar disk was a favorable environment for this process to occur in the terrestrial planet region.

## 2. Compositional constraints

### 2.1 Isotopic composition

Mass-independent isotopic anomalies of nucleosynthetic origin are the most reliable indicators of the provenance of the material incorporated into a planet because they are not easily modified by chemical reactions or physical processes such as evaporation and condensation. The isotope anomalies are typically reported in the µ-notation as the parts per million deviations from the terrestrial standard. In a broad sense, Solar System bodies can be subdivided into two groups based on their isotopic anomalies, dubbed the carbonaceous (CC) and non-carbonaceous (NC) groups. This fact is known as the isotopic dichotomy (Warren, 2011; Budde et al., 2016). Carbonaceous chondrites belong to the CC group, while ordinary and enstatite chondrites belong to the NC group. All chondrites derive from planetesimals that formed relatively late (~2-4 My after CAI) in the disk (after $^{26}$Al was mostly extinct). But, importantly, the NC-CC dichotomy is also already evident for iron meteorites and achondrites (Budde et al., 2016; Kruijer et al., 2017), which all derive from planetesimals formed earlier, probably within the first My of Solar System history (Spitzer et al., 2021). This implies that the isotopic dichotomy was established early in the protoplanetary disk and persisted over the disk's lifetime. Thus, at any time, NC and CC objects had to form at different locations from materials that did not experience significant radial mixing. Because CC chondrites are rich in water and other volatile elements, it is generally accepted that they formed farther from the Sun than NC meteorites. Thus, CC material is characteristic of the composition of the outer disk and NC material of that of the inner disk.

The origin of the isotopic dichotomy is debated (see discussion in Kleine et al., 2020), but for the purpose of this work it does not matter. What is important is that, within the pebble accretion scenario, the radial drift of dust from the outer to the inner disk would eventually bring CC material into the inner Solar System, modifying the isotopic compositions of indigenous NC bodies towards more CC-rich compositions. Thus, if the terrestrial planets of the Solar System formed by pebble accretion, their isotopic compositions should have become increasingly CC-rich as they grew, resulting in a larger CC fraction in more massive, or later-formed objects.

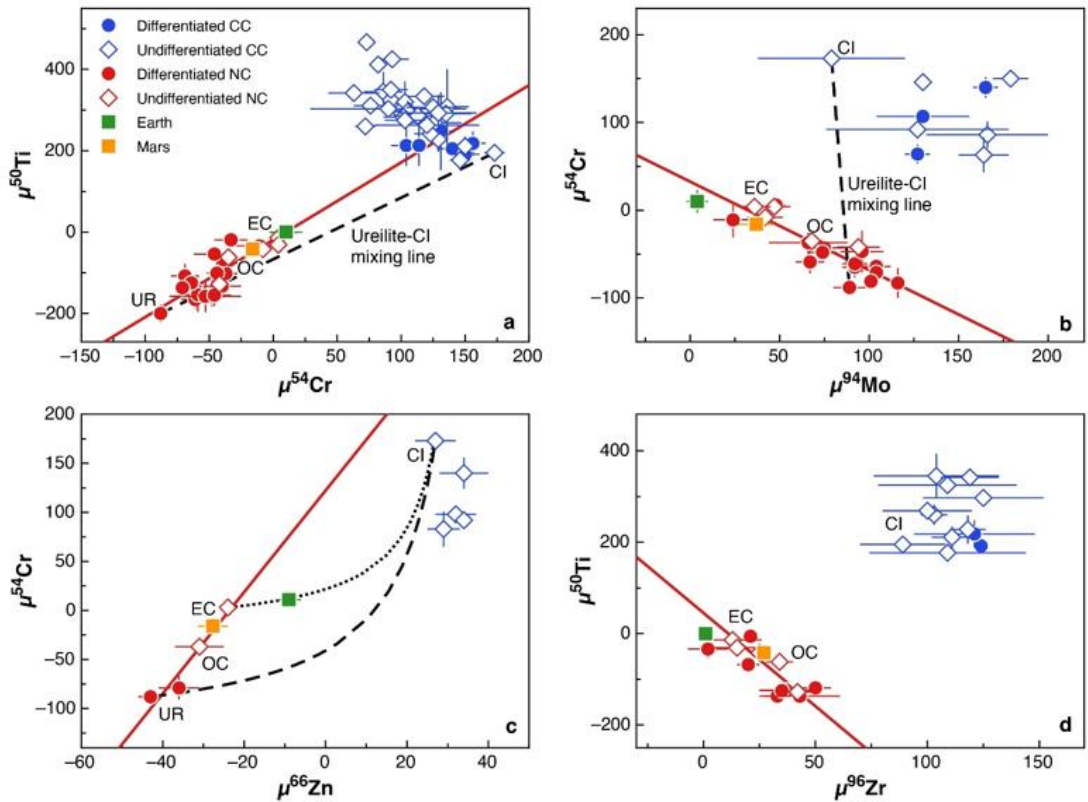

Fig 1. NC (red) and CC (blue) meteorites, together with Earth (green) and Mars (orange) in plots of $\mu^{54}$Cr vs. $\mu^{50}$Ti (a), $\mu^{94}$Mo vs. $\mu^{54}$Cr (b), $\mu^{66}$Zn vs. $\mu^{54}$Cr (c), and $\mu^{96}$Zr vs. $\mu^{50}$Ti (d). The red line is an empirical fit of the distribution of all NC meteorites (Earth and Mars excluded) and the black dashed line is a mixing line between ureilites (an NC achondrite labelled UR) and CI (a CC chondrite); such mixing has been suggested by Schiller et al. (2018) to account for the isotopic compositions of Earth and Mars. The dotted line in panel (c) is a mixing line between Earth and CI chondrites, for which the Cr and Zn abundances and isotopic anomalies are known, extrapolated to the NC trend (red line) by subtracting an increasing CI component from Earth; note that this line intersects the NC trend at the isotopic composition of enstatite chondrites (EC) and that no a priori knowledge of the Zn concentration in the NC end-member is required, because it is determined by the known Zn concentrations of the CC end-member and the BSE. Instead, the mixing line connecting CI chondrites and ureilites does not pass through the composition of the Earth or Mars. OC = ordinary chondrites.

Schiller et al. (2018) argued that this is exactly what is observed in the data. These authors found that the $\mu^{48}$Ca values of meteorites progressively increase from some NC achondrites (the ureilites, which have the most negative $\mu^{48}$Ca) to NC chondrites, Mars and Earth (the latter with $\mu^{48}$Ca=0 by definition). Because CC chondrites have positive $\mu^{48}$Ca values, Schiller et al. (2018) concluded that this trend reflects the progressive penetration of CC dust into the inner Solar System and its incorporation into bodies formed there. Onyett et al. (2023), based on nucleosynthetic variations of $\mu^{30}$Si arrived at a similar conclusion, although their data are questioned in Dauphas et al. (2024). These arguments can be extended to other lithophile and non-volatile Fe-group elements like Cr and Ti, because the isotope anomalies among Ca, Cr, and Ti are correlated in both the NC and CC reservoirs. Figure 1a shows the composition of meteorites, Earth, and Mars in Cr and Ti isotope space (note that the use of $\mu^{48}$Ca instead of $\mu^{54}$Cr or $\mu^{50}$Ti would give an analogous portrait). This plot reveals that as proposed by Schiller et al. (2018), Earth could have accreted 60% of its mass from material having an ureilite-like isotopic composition and 40% from CC material[1]. This appears perfectly consistent with an influx of CC dust into the inner Solar System and with Mars and Earth drawing a large fraction of their mass from this influx, as expected if these planets grew predominantly by pebble accretion. However, if the NC end-member were taken to be isotopically identical to the enstatite chondrites (EC), then the inferred fraction of CC material in the Earth would be much smaller, roughly 5% (Warren, 2011; Dauphas et al., 2024; Burkhardt et al., 2021).

Clearly, distinguishing between these very different interpretations of the same data requires additional constraints. Below we will show that these constraints are provided by considering the largest possible set of isotope anomalies and a more diverse set of meteorites, which cover the entire span of planetesimal formation times in both the NC and CC reservoirs. Using such a global data set, we assess three key predictions of the pebble accretion model, namely (*i*) that the isotopic composition of NC reservoir evolved over time due to the progressive addition of CC dust, (*ii*) that the terrestrial planets accreted a large fraction of CC material, and (*iii*) that CC material dominated the later stages of accretion.

### 2.1.1 Did the NC isotopic composition evolve over time?

As noted above, Schiller et al. (2018) argued that the isotopic composition of the NC reservoir evolved over time through the addition of CC dust. However, two observations indicate that this did not happen. First, Earth, enstatite chondrites, and aubrites almost perfectly share the same isotope anomalies. This is problematic for the pebble accretion paradigm where Earth accretes material continuously and therefore its NC-CC mixture reflects the integral of the contamination of the inner disk by CC dust over the disk's lifetime; by contrast, the enstatite chondrite parent bodies should have formed at a given point in time by the streaming instability and should, therefore, reflect a temporal snapshot of the disk's composition at a given radial distance. For a continuous function (i.e., the fraction of CC material in the inner disk over time), the integral must be equal to the value of the function in at least one point, so the match could just be coincidental. However, aubrites, which are NC achondrites and as such formed approximately 1 My before the enstatite chondrites (Sugiura and Fujiya, 2014), also have similar isotopic anomalies as enstatite chondrites and the Earth. So, the coincidence would have to repeat twice at two radically different times. This is inconsistent with the expectation of a monotonic increase of the CC fraction in the inner disk due to the radial drift of outer Solar System dust.

---

[1] Schiller et al., 2018 proposed that Earth accreted from ureilites and CI chondrites, but the mixing line would be a bit offset. Accretion from a mixture of CC chondrites does not change the essence of their argument.

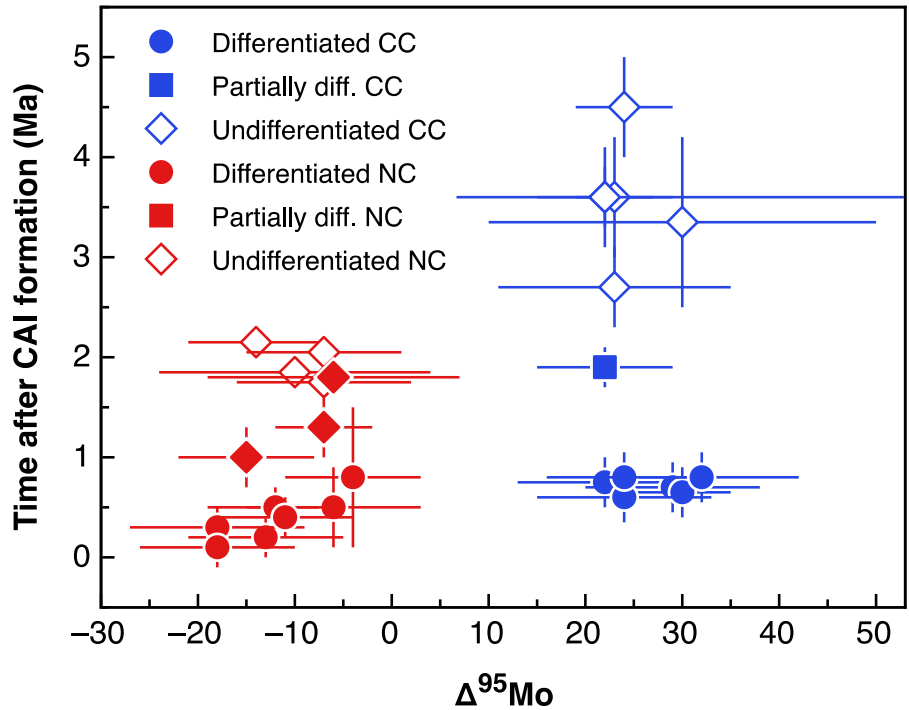

Fig 2. Δ$^{95}$Mo of NC (red) and CC (blue) meteorites versus the inferred accretion ages of their parent bodies. Δ$^{95}$Mo is defined as the deviation of a sample's Mo isotopic composition from a theoretical *s*-process mixing line passing through the origin in the μ$^{95}$Mo-μ$^{94}$Mo plot (Budde et al., 2019). The NC and CC reservoirs are each characterized by distinct Δ$^{95}$Mo values, which in each reservoir remained approximately constant over time, demonstrating that there has been no influx of CC material with elevated Δ$^{95}$Mo values into the NC reservoir (i.e., the inner disk). Modified from Kleine et al. (2020). Accretion ages from Sugiura and Fujiya (2014), Desch et al. (2018), and Kleine et al. (2020).

Second, in more general terms, there is no temporal trend in the isotopic compositions of either the NC or the CC reservoir. This can be clearly seen when the isotopic signatures of a comprehensive set of meteorites, including undifferentiated and differentiated meteorites from both the NC and CC reservoirs, are plotted against the accretion time of their parent bodies. Figure 2 shows the characteristic Mo isotope composition of the NC and CC reservoirs (expressed as Δ$^{95}$Mo, defined in the figure caption) plotted against the accretion ages of meteorite parent bodies from both groups. The advantage of using Δ$^{95}$Mo is that the NC and CC reservoirs are characterized by single but distinct Δ$^{95}$Mo values with little to no variations within each reservoir (Budde et al., 2019). Thus, the near-constant Δ$^{95}$Mo value of the NC reservoir demonstrates that there has been no significant addition of CC material having higher Δ$^{95}$Mo into the NC reservoir.

### 2.1.2 How much CC material was accreted by Earth and Mars?

As mentioned above, the CC fractions inferred for Earth and Mars using variations in μ$^{48}$Ca, μ$^{54}$Cr, or μ$^{50}$Ti can be very high or very low, depending on which meteorite group is taken to be the NC end-member in these bodies. However, a plot of $\mu^{66}$Zn versus $\mu^{54}$Cr (Fig. 1c) shows that the $\mu^{54}$Cr value of the NC end-member dominating the Earth's composition must be EC-like and not related to the ureilites. The Zn isotope data also indicate that Earth contains about 6 wt.% CI chondrite-like material (Steller et al., 2022; Savage et al., 2022), while Mars shows no detectable CC fraction (Kleine et al. 2023; Paquet et al., 2023). Thus, Zn isotopes provide powerful evidence against the proposal that the isotopic compositions of Earth and Mars reflect the progressive addition of CC dust to an NC seed having an ureilite-like isotopic composition.

Further evidence for an overall low CC fraction in Earth and Mars comes from the isotope anomalies in elements produced in the slow neutron capture process (*s*-process), such as Mo, Zr, Nd, and Ru. Fig. 1b and 1d show that in the $\mu^{54}$Cr vs. $\mu^{94}$Mo and $\mu^{50}$Ti vs. $\mu^{96}$Zr plots, NC meteorites, Mars, and Earth plot along a line that does not point towards the CC group. The mixing line between ureilites and CI chondrites completely misses the position of Earth in the $\mu^{54}$Cr vs. $\mu^{94}$Mo diagram, as well as any line connecting ureilites to an arbitrary mixture of CC meteorites. Similarly, in the $\mu^{50}$Ti vs. $\mu^{96}$Zr plot any mixing line between NC and CC meteorites would not pass through the composition of Earth (note that there are no $^{96}$Zr data for ureilites). Burkhardt et al. (2021) showed that the same trend is also seen for $^{145}$Nd and $^{100}$Ru. Thus, for the *s*-process elements, Earth is not intermediate between NC and CC meteorites, but is an end-member on the NC trend, which does not point towards the CC field. This observation demonstrates that, compared to meteorites, Earth is enriched in *s*-process matter (Burkhardt et al., 2011; Render et al., 2022) and excludes the incorporation of a large fraction of CC material. From a quantitative analysis, Burkhardt et al. (2021) concluded that the most probable CC contribution to the bulk Earth is only 4% and that a CC fraction of more than 20% is excluded at the 97.5% confidence level. This is consistent with the aforementioned ~6 wt.% CC contribution estimated based on the $\mu^{66}$Zn-$\mu^{54}$Cr systematics (Fig. 1c).

An alternative explanation for the terrestrial offset from the ureilite-CC mixing line for *s*-process elements has been proposed by Onyett et al. (2023), as follows. The products of the *s*-process are typically carried by silicon-carbide (SiC) interstellar grains, which are highly refractory, while the species which have an isotopic composition complementary to SiC and which, when combined with the SiC grains, yield the assumed CI isotopic composition of the pebbles, are more volatile. On this basis it has been suggested that as pebbles penetrate through the Earth's primitive atmosphere, these more volatile species evaporate and are partially recycled back into the disk (Brouwers and Ormel, 2020), while SiC grains remain in the solid, refractory component of the pebble and are preferentially accreted by the planet. This process could indeed offset Earth from a ureilite-CC mixing line. But Fig. 1 (b, d) shows that Earth is not anomalous: the whole NC population follows a linear trend transverse to any ureilite-CC mixing line. Note that NC meteorites come from planetesimals, bodies presumed to be lacking any atmosphere, for which the invoked SiC preferential accretion mechanism would not apply. Moreover, the NC trend is defined not only by *s*-process elements but extends to the Fe-group elements as well, and so cannot reflect the preferential enrichment in solely *s*-process carriers.

In summary, three sets of isotope data have been used to infer the amount of CC material accreted by Earth and Mars, namely (*i*) non-volatile Fe-group elements (e.g., Ca, Cr, Ti), (*ii*) the moderately volatile element Zn, and (*iii*) *s*-process elements (e.g., Zr, Mo). Of these, only the first group of elements may be used to infer a high CC fraction, but, importantly, does not require it. By contrast, all three groups of elements consistently indicate a low CC fraction in Earth and Mars.

### 2.1.3 When did the Earth accrete CC material?

While the CC fraction in the bulk Earth is small, the CC contribution recorded in the isotopic signature of siderophile elements in a planet's mantle can in principle be higher. This is because the budget of siderophile elements in the mantle is dominated by the material added towards the end of accretion, while most of the siderophile elements from earlier accretion stages reside in the core (Dauphas, 2017).

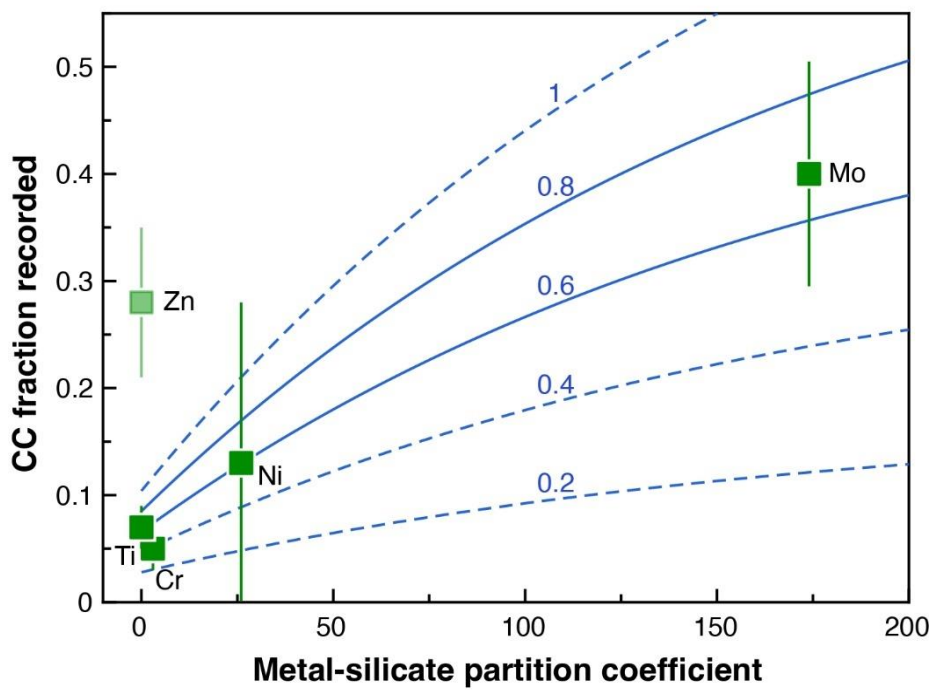

Fig 3. Fraction of CC material versus metal-silicate partition coefficient for different elements. The CC fractions are calculated using the difference in isotopic composition of the BSE compared to those of enstatite chondrites and CI chondrites. The metal-silicate partition coefficients are calculated from the BSE abundances and assuming a single stage of metal segregation. Blue lines (dashed and solid) are results of simple two-step accretion model in which the first 90% of Earth accretion were 99% NC, while the last 10% consisted of a NC-CC mixture, with the CC fraction reported in the label. The best fit to the data is obtained by assuming that the CC fraction in the last 10% of accretion was 60-80% (solid lines). Zinc plots above the model lines because the late-added CC material was volatile-rich compared to the earlier-accreted NC material.

Schiller et al. (2020) found that the BSE has the same $\mu^{54}$Fe value as CI chondrites, which itself is distinct from all other meteorites. This observation appears consistent with the idea that Earth accreted roughly the second half of its mass from material akin to CI chondrites, which are presumed to represent the composition of pebbles from the outer disk. However, Hopp et al. (2022) have shown that in this case the BSE should also have a CI chondrite-like isotope composition for Ni (another siderophile element with similar siderophility as Fe), yet CI chondrites have the largest $^{64}$Ni anomaly (compared to the BSE) among all meteorites (Spitzer et al., 2024). In general, the use of $\mu^{54}$Fe to trace the provenance of Earth's accreted material is difficult because the NC and CC domains overlap for $\mu^{54}$Fe, unlike for Ni isotopes (see Fig. 8 in Hopp et al., 2022, for a $\mu^{54}$Fe vs. $\mu^{64}$Ni diagram).

A more reliable constraint on the amount of CC material accreted during the late stages of Earth's growth is provided by Mo. For this element, the CC fraction can be deduced from the $\Delta^{95}$Mo quantity (see Fig. 2 for a definition), which is independent of the internal isotope variations within each reservoir. On this basis, a CC contribution to the BSE' Mo budget of about 30-60% has been determined (Budde et al. 2019, Burkhardt et al., 2021). Figure 3 shows the CC fraction recorded in the BSE for elements with different metal-silicate partition coefficients, compared with a model in which the first 90% of mass delivered is 99% NC and the final 10% is an NC-CC mix with a CC fraction of 60-80%. The high CC fraction recorded by volatile Zn can be explained if the late-delivered material is enriched in Zn by a factor of about 8 compared to the early-delivered material. Thus, the late-delivered material appears to be enriched in both CC and volatile elements.

This conclusion may seem to give a second chance to the pebble accretion paradigm, because it may reflect the late arrival of CC dust in the inner disk. Indeed, in this case, Earth would have accreted CC material in large abundance only at the very end of the disk's lifetime, allowing the bulk Earth to remain isotopically similar to enstatite chondrites. Given the radial drift speed of dust in a disk, this scenario requires that the original boundary between the NC and CC parts of the disk was at about 100 astronomical units (au) (Burkhardt et al., 2021). Even the accretion of ~40% of the bulk Earth's mass from CC material advocated by Schiller et al. (2018) requires that the CC dust penetrated the inner Solar System not before 3.8 My, and that the NC-CC boundary was beyond 30 au from 0.2 to 3 My (see Fig. S2 in Johansen et al., 2021), i.e. within the Kuiper Belt. This can be justified by some disk evolution models (Liu et al., 2022), but is at odds with the physical characteristics of Kuiper Belt Objects (KBOs). Regardless of their dynamical sub-classes, KBOs with diameter smaller than 1,000 km have densities smaller than 1.5 g/cm$^3$, indicating a high porosity and formation after $^{26}$Al was extinct (i.e. after 4 My; Brown 2013, Bierson & Nimmo 2019). As such, these bodies did not undergo metal-silicate differentiation, yet the parent bodies of CC iron meteorites (which accreted at ~1My: Spitzer et al., 2021) are definitively differentiated. It is therefore hard to argue that these bodies formed within the Kuiper Belt, which is what the pebble accretion model would require. Moreover, both enstatite chondrites and aubrites accreted before 3.8 My (i.e., before the assumed CC arrival time), and so according to Johansen et al. (2021) they should not have incorporated any CC material. However, aubrites and enstatite chondrites are isotopically similar to Earth which, in the context of the pebble accretion model of Schiller et al. (2018), would imply they contain ~40% CC material.

In summary, when the full set of elements and the entire suite of meteorites is considered, isotopic considerations contradict the pebble accretion hypothesis. This model predicts that the NC isotopic composition evolved over time due to the progressive addition of CC dust, and that the terrestrial planets accreted a large fraction of CC material. However, both predictions appear inconsistent with the isotopic data. A possible solution to this problem could be to assume that the NC-CC barrier was far out in the disk, so that CC dust arrived in the inner disk only late, after the accretion of NC meteorite parent bodies was complete. However, in this case, the close isotopic match of enstatite chondrites and aubrites (both of which formed prior to the assumed arrival of CC dust) with the Earth cannot explained. Also, the existence of differentiated CC planetesimals would be difficult to reconcile with the lack of differentiation inferred for Kuiper Belt objects. The only way out of this paradox is that a structure in the disk – often called the *Jupiter barrier* (Kruijer et al. 2017), but not necessarily linked to the formation of Jupiter (Brasser and Mojtzsis, 2020) – blocked the drift of CC dust into the inner Solar System. Indeed, the analysis of CAIs in ordinary chondrites (Haugbolle et al., 2019) shows that only particles smaller than 200μm have been accreted from the outer disk; these particles appear to have carried a negligible fraction of the total mass because they did not shift these meteorites' isotopic compositions off the NC line, towards the CC compositional field (Fig. 1). Consequently, if the terrestrial planets grew by pebble accretion, they did so by accreting predominantly local dust, itself trapped in a disk structure (i.e. in a dust ring).

In the classic model, the accretion of Earth from a local population of planetesimals produces a planet that is predominantly NC in a natural way. The CC contamination is due to objects scattered towards the inner Solar System during the period of mass-growth of the giant planets (Raymond and Izidoro, 2017). The large CC fraction of the final part of Earth's accretion, and the absence of an equivalent signature for Mars, imply that the mass budget of the CC material was dominated by stochastic collisions with a very small number of CC planetary embryos (Nimmo et al., 2024).

### 2.2 Chemical composition

The chemical composition of the bulk silicate Earth is characterized by a gradual volatile depletion of lithophile elements with a condensation temperature below 1,400 K. This is in contrast with other differentiated but smaller bodies, like Vesta (parent body of HED meteorites), which shows a step-function depletion pattern (Fig. 4).

In the classic scenario, the BSE's volatile depletion pattern is consistent with the accretion of an early generation of planetesimals, each characterized by a step-function volatile depletion pattern like Vesta, but with a step at different temperatures, depending on the conditions of these bodies' accretion and temperature evolution (Sossi et al., 2022). In this model, chondrites (i.e. late-formed planetesimals) are not a major constituent of Earth. By contrast, it seems more problematic to explain the BSE's volatile depletion pattern in the framework of the pebble accretion hypothesis, at least in the absence of a proto-Earth atmosphere (see below for this case). Pebbles are small objects, at thermal equilibrium with the disk. Thus, Earth should accrete from pebbles only the elements whose condensation temperature is higher than the temperature of the disk where Earth grows. The disk cools over time, but in the pebble accretion model the growth of a planet is close to exponential (given that the accretion rate is proportional to a planet's mass to a power close to unity); thus, most of the mass is accreted in a relatively narrow interval of time, during which the temperature does not evolve much. The expected volatile depletion pattern is therefore much steeper than observed for the BSE and should be more similar to the step-function observed for instance for Vesta.

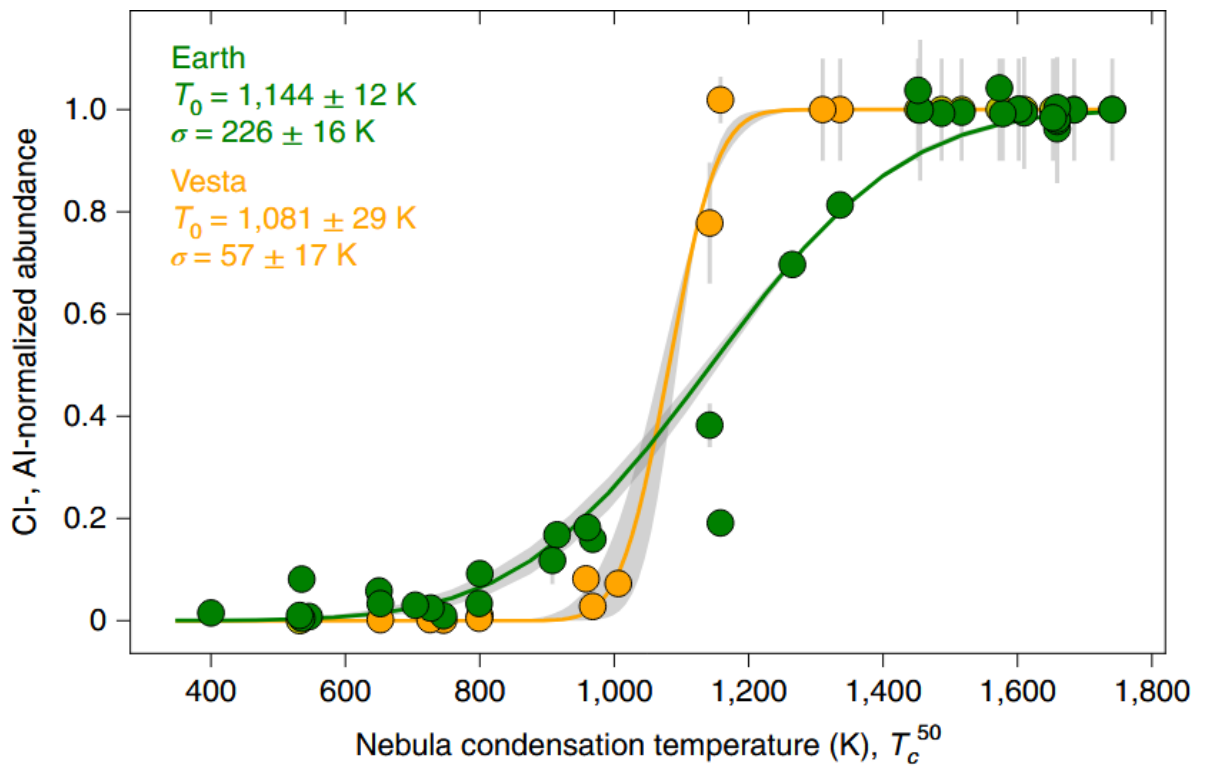

Fig. 4. The abundance of elements as a function of their condensation temperature in the BSE (green) and HED meteorites (orange). From Sossi et al. (2022).

In their pebble accretion model, Johansen et al. (2021) proposed that Earth accreted in a cold disk, at the snowline. So, in principle it should have accreted volatiles in CI chondritic (i.e. solar) proportions. In this picture, Earth was surrounded by a hot primitive atmosphere of H and He. Pebbles entering such an atmosphere underwent evaporation, from more volatiles to more refractory species, as they penetrated deep into the atmosphere (Brouwers and Ormel, 2020). The upper part of the atmosphere was gravitationally unbound to the planet, so the volatiles that evaporated there were not accreted. This model was calibrated to achieve a strong depletion in the water content of Earth, even if the planet accreted at the snowline. However, the temperature at the base of the unbound atmosphere in this model was 400 K or less, meaning that Earth should show no depletion for volatile elements with higher condensation temperatures, unlike what is observed (Sossi et al., 2022). In the more general case, whatever the temperature at the base of the unbound atmosphere, we would expect a resulting volatile depletion described by a step-function, with the step located at that temperature. To our knowledge, no pebble accretion model reproducing the BSE's volatile depletion pattern has ever been published.

### 3. Dynamical constraints

Pebble accretion can operate only as long as there is gas in the disk. Thus, if planets predominantly accrete by pebble accretion, necessarily they form within the lifetime of the disk. We will examine the chronological constraints in the next section and here focus on the fact that planets above a few Mars masses embedded in a disk of gas should undergo significant ("type 1") radial migration. For the mass interval relevant for terrestrial planets, the migration speed is proportional to the planet's mass. So, migration should have been slow for Mercury, fast for Venus and Earth and slow for Mars. Earth and Venus should have pushed Mercury inwards in a mean motion resonance and left Mars behind. But the regular orbital spacing between the four terrestrial planets (roughly each planet being at a distance from the Sun that is 1.5x that of the previous planet) does not show any evidence that this happened. In the conventional accretion scenario, this problem is solved by assuming that most growth happened after the gas disk had dissipated.

To circumvent this problem in the pebble accretion paradigm, Johansen et al. (2021) postulated that the planets started to form at 1.6 au (the snowline) in a temporal sequence. As a planet grew by pebble accretion it migrated inwards, leaving the place for a new planet to start forming at the same location. This produced a definite mass vs. semi major axis distribution of the planets at the end of the disk lifetime: Venus was the innermost and most massive planet, then came an at-the-time sub-venusian Earth, then Theia, then Mars. Theia and the proto-Earth later collided to form the actual Earth, more massive than Venus. This model solves the problem of leaving Mars too much behind during migration: Mars barely migrated by 0.1 AU, but Earth and Theia migrated by 0.6 and 0.3 au respectively, Venus by 0.9 au, thus reproducing the current separations. Obviously, Mercury does not fit in this pattern (the innermost planet should have formed first and be the most massive), so it is invoked that Mercury formed by a distinct process (Johansen and Dorn, 2022). Mars in this model, being the planet that forms the last, should be the richest in CC material, in direct contradiction with the Zn isotope analyses that detect no CC contribution to the composition of this planet (Kleine et al., 2023; Paquet et al., 2023) and with the results of Schiller et al. (2018), who inferred a larger CC fraction in Earth than in Mars.

There is another possibility to solve the migration problem. Modern models of terrestrial planet accretion from a ring of planetesimals invoke that the gas surface density of the disk peaked at 1 au (Broz et al., 2021; Woo et al., 2023), possibly due to the action of magnetized disk winds (Ogihara et al., 2015). Such a disk would have exerted convergent migration at ~1 au also in the case planets formed by pebble accretion. Because the dust also tends to concentrate near the pressure maximum, the mass distribution of the terrestrial planets, with small Mercury and Mars at the extremes and Venus, Earth and Theia at the center, is likely to be produced. Convergent migration should have brought the five planets into a resonant chain, but the latter would have been presumably broken when the giant planets became temporarily unstable (Tsiganis et al., 2005), sometime after the removal of the gas.

The problem here is that pebble accretion is an oligarchic process (i.e. the relative mass accretion rate $\frac{1}{M}\frac{dM}{dt}$ typically decreases as $M$ increases; Ida et al., 2016). Thus, in this scenario of contemporary formation of the planets near a maximum of the gas density distribution (rather than the sequential formation of Johansen et al., 2021), multiple planets should grow with comparable masses. A too large number of protoplanets is unstable, so they tend to eliminate each other by mutual scattering or mutual collisions (Levison et al., 2015). In the terrestrial planet region, mutual collisions would be the most natural outcome, because the orbital velocities are much larger than the escape velocities from the protoplanets. Hence, the most natural outcome of pebble accretion is the formation of a set of planetary embryos, which would then mutually collide to form the terrestrial planets (Lambrechts et al., 2019). This growth mode is dynamically plausible and would be a hybrid between the pebble accretion model and the classic model of formation of terrestrial planets via giant impacts. But it would not be correct to say that Earth formed predominantly by pebble accretion.

To avoid the formation of too many planetary embryos and the occurrence of an intense giant impact phase, one must postulate that only five pebble-accreting planetary seeds (say objects larger than Ceres – 1,000km in diameter) formed in the inner Solar System, and no substantial population of planetesimals. This is because, if a substantial population of planetesimals existed, inevitably the mutual planetesimal collisions would have produced numerous pebble-accreting seeds (Woo et al., 2023).

But the absence of a substantial population of planetesimals is inconsistent with the high bombardment rate witnessed by the old surfaces of Mars and the Moon (Neukum et al., 2001). Nesvorny et al. (2023) showed that the lunar crater record requires that 4.35 Gy ago (when the record started), the population of Earth-crossing planetesimals comprised ~1,000 objects with diameter D>100km and ~35,000 objects with D>10km. It is dynamically implausible that such a population suddenly appeared as fragments of a single (or a few) larger bodies. In addition, as large bodies are likely differentiated, the lunar projectiles would be expected to be achondritic, inconsistent with the detection of a substantial chondritic component in the lunar bombardment record at the time of late basin formation (Joy et al., 2012). Thus, the lunar projectiles must be the survivors of a much larger population of planetesimals, accounting originally (i.e. at the end of the gas-disk phase) for more than 2x10⁵ bodies with D>100km. This renders the existence of only five pebble-accreting Ceres-size seeds quite implausible.

A general problem of pebble accretion inward of the snowline is that rocky pebbles should be small and strongly coupled with the gas, rendering pebble accretion inefficient (Batygin and Morbidelli, 2022). In fact, laboratory experiments show that the fragmentation velocity of rocky grains is 1m/s, about one order of magnitude lower than that of icy grains (Blum and Munch, 1993). Thus, pebbles can grow only until their dispersion velocity in the turbulent disk equals 1m/s. The dispersion velocity is $v_{turb} = \sqrt{\alpha\tau}c_s$ (Birnstiel et al., 2009), where $\alpha$ is the turbulence strength parameter (Shakura and Sunyaev, 1973), $\tau$ is the pebble Stokes number and $c_s$ is the sound speed in the disk. This formula reveals that the pebble Stokes number is proportional to the square of the fragmentation velocity, and thus should be 100x smaller for rocky than for icy pebbles (or even smaller, given that $c_s$ scales as $\sqrt{T}$, where $T$ is the temperature).

The pebble Stokes number is key to determine the efficiency of pebble accretion for two reasons. First, the collision parameter for pebble accretion is $b = (\frac{\tau}{0.1})^{1/3}R_H$, where $R_H$ is the Hill radius of the protoplanet (Ida et al., 2016). Second, the thickness of the pebble layer is $H_p = \sqrt{\alpha/\tau}\,H_g$ (Chiang and Youdin, 2010), where $H_g$ is the scale height of the disk of gas. The value of $\tau$ therefore dictates whether $H_p > b$ or $H_p < b$, i.e. whether pebble accretion occurs in the 3D or 2D regime. Accretion in the 3D regime is very inefficient, being proportional to $\rho_p b^3 \Omega$, where $\rho_p$ is the volume density of the pebble layer near the disk's midplane and $\Omega$ is the orbital frequency. Only the 2D regime, at a rate proportional to $\Sigma_p b^2 \Omega$, where $\Sigma_p$ is the surface density of the pebble layer, is considered relevant (Ormel, 2017).

The magenta curve in Fig. 5 shows the protoplanet mass at which accretion transitions from the 3D to the 2D regimes, as a function of the temperature of the disk at 1 au. It assumes $\alpha = 10^{-4}$, which is often considered to be a lower limit for the intrinsic turbulence in a disk. In this case, protoplanets must exceed the mass of Mars, or even of Earth, to accrete in the 2D regime, implying that pebble accretion should have played no role, even for the formation of planetary embryos. The green curve, instead, assumes no intrinsic turbulence, so that $\alpha$ is regulated by the streaming instability (Li and Youdin, 2021). In this case, the threshold mass is comparable to the mass of Ceres. The streaming instability is unlikely to produce directly objects of such masses (Johansen et al., 2015 – their Fig. 4; Klahr and Schreiber, 2020). Thus, these objects would need to form by mutual collisions of a planetesimal population. Pebble accretion and planetesimal accretion would then compete in the further growth of these objects, but most likely forming a set of planetary embryos, rather than just five terrestrial planets (including Theia). We stress that in this hybrid scenario the pebble component still needs to be "local" to satisfy the isotopic constraints discussed in Sect. 2.1.

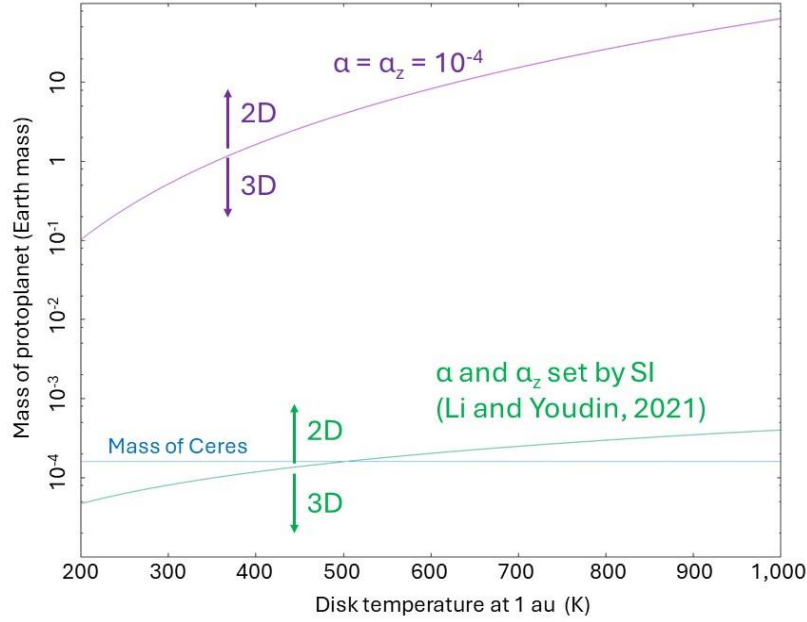

Fig. 5. Mass threshold for the transition from 3D to 2D pebble accretion, as a function of the disk temperature at 1 au. Here we use the formulae: $H_g = 0.045 \sqrt{\frac{T}{500K}}$ au, $c_s = 1343 \sqrt{\frac{T}{500K}}$ m/s (Bitsch et al., 2015), $\tau = (\frac{1m/s}{c_s})^2/\alpha$, $H_p = \sqrt{\alpha_z/\tau} H_g$, $b = (\frac{\tau}{0.1})^{1/3} R_H$ and $R_H = 1$ au $(\frac{M_p}{3M_*})^{1/3}$ and impose $H_p = b$. The magenta curve assumes $\alpha = \alpha_z = 10^{-4}$. The green curve assumes that $\alpha = 10^{-4}(\frac{\tau}{0.1})^2$ and $\alpha_z = \alpha/5$, as resulting from streaming instability simulations in a disk without intrinsic turbulence (Li and Youdin, 2021). The blue horizontal line denotes the mass of Ceres, for reference. Thus, pebble accretion in the 3D regime is very inefficient (see text).

We also remark that these problems disappear for pebble accretion beyond the snowline, because the Stokes number $\tau$ is expected to be 100x larger. As the formulae in the caption of Fig. 5 reveal, the threshold mass for the transition from the 3D to the 2D regime scales as $1/\tau^{5/2}$ and, at equal planetary mass, the pebble accretion rate in the 2D regime scales as $\tau^{2/3}$. This is why giant planet cores could grow beyond the snowline within the lifetime of the disk (Morbidelli et al., 2015).

## 4. Chronological constraints

Planets growing by pebble accretion would complete their growth within the lifetime of the disk (about 5 My according to paleomagnetic measurements; Weiss et al., 2021). As they grow, the planets should melt due to the released potential energy and $^{26}$Al decay. In these molten bodies, there should be a continuous segregation of metal and silicate, while both are being delivered by the accreted pebbles (Olson et al. 2022). This scenario of continuous core formation has been investigated in detail for application of the short-lived Hf-W system to date core formation. $^{182}$Hf is a radioactive lithophile element that decays into the siderophile element $^{182}$W with a half-life of 8.9 My. The faster the planet grows to completion, the more radiogenic is the final mantle in $^{182}$W. For a planet accreting within 5 My, the final value of $\varepsilon^{182}$W would be between 10 and 20, depending on the exact growth history of the planet (Jacobsen, 2005; Kleine et al., 2009), whereas the observed value for the BSE is $\varepsilon^{182}$W =1.9, where $\varepsilon^{182}$W =0 is the chondritic reference.

The much lower $\varepsilon^{182}$W of the BSE compared to the expected value for pebble accretion might be explained by admitting that Earth did not form solely by pebble accretion, but suffered one late-stage giant impact, as is commonly invoked for the formation of the Moon. In this scenario, pebble accretion would have formed a proto-Earth and Theia within the lifetime of the disk, which then collided at a much later time. The metal-silicate equilibration during the impact event would have substantially lowered the value of $\varepsilon^{182}$W, making it potentially compatible with the measured value (Yu & Jacobsen 2011, Johansen et al., 2023).

The degree to which a late collision can reduce the pre-existing $^{182}$W anomaly depends on the size of the impactor and the effective equilibration factor $k$. The latter depends both on how much of the impactor core interacts with the target mantle, and what fraction of the target mantle interacts with the core (Deguen et al. 2011).

Figure 6 illustrates what combinations of $k$ and impactor/target mass ratio are compatible with reducing an initial $\varepsilon^{182}$W to the present-day terrestrial value. In the pebble accretion scenario, the rapid early formation of Earth results in strongly positive $\varepsilon^{182}$W values (of the order of 10-20). The figure shows that, to reach the present BSE value of $\varepsilon^{182}$W from an original value of 12, the Theia/proto-Earth mass ratio should have been ~0.3 for $k$=1; even a mass ratio of 1 would still require $k$>0.4 (see Appendix for a description of how the values reported in the figure have been computed). In the Johansen et al. (2021) scenario, the mass ratio is ~0.5, requiring $k$~0.7. Johansen et al. (2023) find a similar result, requiring $k$>0.5. If the original $\varepsilon^{182}$W was 18, this would have required even larger mass ratio and equilibration factor.

However, such large equilibration factors are inconsistent with two other observations. First, to reconcile the BSE's Ni isotope composition with the pebble accretion model, Onyett et al. (2023) argued for equilibration factors of only 0.1 – 0.3, which are too low to also account for the BSE's $^{182}$W composition (Fig. 6). Second, the high degree of equilibration needed to account for the BSE's $^{182}$W composition is also inconsistent with our current knowledge of the likely $k$ based on fluid dynamics. Landeau et al. (2021) calculated the effective $k$ for three impact scenarios: a canonical Moon-forming impact (diamond in Fig. 6); an impact onto a fast-spinning Earth (square), and a 100 km radius impact (circle). The small impact results in almost perfect equilibration, as expected. However, the two large impacts result in small equilibration factors ($k$<0.4), consistent with previous work (Dahl & Stevenson 2010, Deguen et al. 2014). These impacts could reduce the Earth's tungsten anomaly to the current value of 1.9 only if the pre-impact anomaly was 2.5 (green dashed line in Fig. 6), but a value of $\varepsilon^{182}W=2.5$ is well below that expected (10-20) for a proto-Earth forming in less than 5 My.

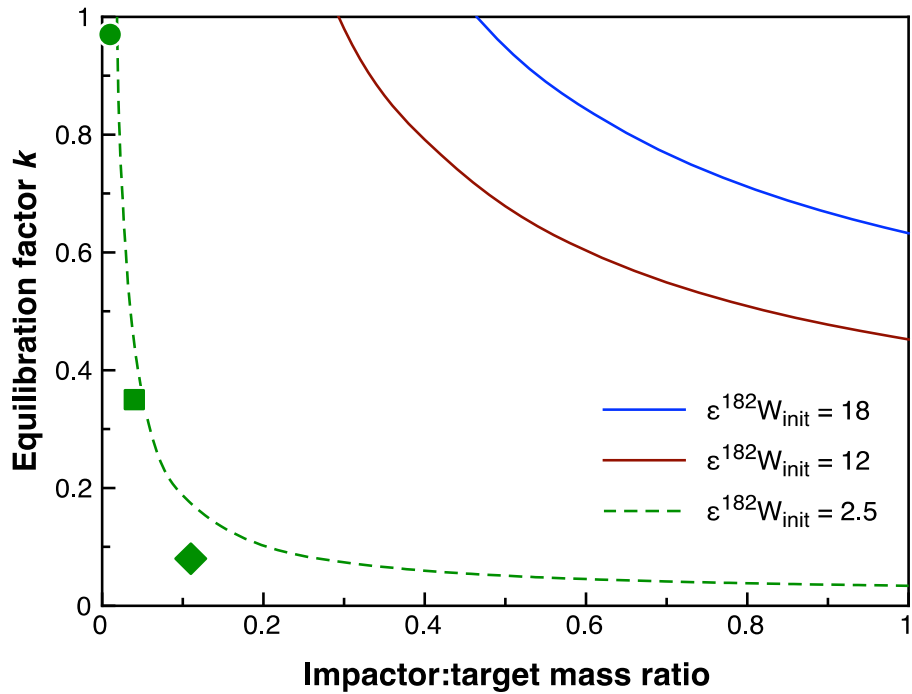

Fig. 6. Combinations of Theia/proto-Earth mass ratio and the equilibration factor *k* required to reduce the BSE's $\varepsilon^{182}W$ from three initial values of 18 (blue curve), 12(red) and 2.5 (green) to its present-day value of $\varepsilon^{182}W=1.9$. The orange symbols are three impact scenarios from Landeau et al. (2021), showing that they could reduce the Earth's tungsten anomaly to the present-day value only if the pre-impact $\varepsilon^{182}W$ was 2.5. The latter is much smaller than $\varepsilon^{182}W = 10 – 20$, expected for a proto-Earth forming within the lifetime of the disk. Here we assume a core mass fraction of 0.33, a partition coefficient of 35 and that both objects are differentiated and have the same initial $\varepsilon^{182}W$. See Appendix.

These results exclude the possibility that the proto-Earth completed its formation within the lifetime of the disk, as expected in the pebble-accretion model, unless it suffered multiple late giant impacts (Olson and Sharp, 2023), which then dominated its mass growth history. In this case, most of the Earth would have grown by giant impacts, as in the classic model of planet formation.

5. **Evidence for multiple giant impacts**

The reference model of terrestrial planet formation by pebble accretion (Johansen et al., 2021) predicts the formation of four planets: Venus, proto-Earth, Theia and Mars (Mercury is assumed to have formed by a different process, as stated in Sect. 3.). Proto-Earth and Theia then collided after gas removal. Thus, Earth would have experienced only one giant impact. Geochemistry arguments, however, suggest that Earth experienced at least two giant impacts, defined as impacts capable of producing a substantial magma ocean and the outgassing of a secondary atmosphere (Tucker and Mukhopadhyay, 2014). The evidence is based on the $^3He/^{22}Ne$ ratio, which is ~10 in the mid-ocean ridge basalts (MORB) basalts, sampling the upper (depleted) mantle, and ~2.5 in the ocean island basalts (OIB) sampling the lower (primitive) mantle. The OIB $^3$He/$^{22}$Ne is the primitive value, reflecting ingassing of a primitive atmosphere, which fractionates the $^3$He/$^{22}$Ne ratio from the nebular value of 1.5 to the OIB value of 2.5 due to the higher solubility of $^3$He in the original magma ocean. The existence of an original magma ocean does not require any giant impact but is simply due to radioactive decay and the blanketing effect of a primitive atmosphere. It would form also in the pebble accretion scenario (Olson et al. 2022). Increasing the $^3$He/$^{22}$Ne ratio beyond 2.5 requires that (*i*) the primitive atmosphere is lost after the removal of the gas from the protoplanetary disk and (*ii*) a new magma ocean forms due to a giant impact, allowing mantle outgassing and the formation of a secondary atmosphere. However, the increase in $^3$He/$^{22}$Ne is buffered by the build-up of the atmosphere and saturates at about ~5. Increasing the $^3$He/$^{22}$Ne ratio beyond this value requires that the atmosphere is lost again, presumably by a multitude of impacts (Schlichting et al., 2015), then the formation of a new magma ocean due to another giant impact that allows a new outgassing episode (Tucker and Mukhopadhyay, 2014). Thus at least two giant impacts onto Earth should have happened after the removal of the disk. This suggests that multiple protoplanets, not just five, should have emerged at the end of the protoplanetary disk phase.

We remark that giant impacts probably also affected other terrestrial planets. A giant impact is usually invoked to explain the large core mass fraction of Mercury (Benz et al., 1988; Asphaug and Reufer, 2014), and the impact of a 1,600-2,700km body is invoked to explain the Martian surface dichotomy (Marinova et al., 2008).

All these considerations strongly suggest that multiple planetary embryos formed in the inner Solar System, which then assembled into the terrestrial planets via a series of mutual giant impacts. As such, pebble accretion cannot have grown the terrestrial planets to completion. It might only have played a role in the formation of planetary embryos, provided that the disk's low viscosity conditions described in Sect. 3 were fulfilled and that the accreted dust was local.

## 6. Conclusions

We have summarized a series of constraints and considerations that argue against a simple model of terrestrial planet formation by pebble accretion. In this model, a substantial flux of pebbles from the outer protoplanetary disk feeds the growth of four objects: Venus, the proto-Earth, Theia and Mars (while Mercury formed at the inner edge of the disk by a different process – Johansen et al., 2021; Johansen and Dorn, 2022). The planets first accrete local material, analogous to ureilites in isotopic composition, then from dust of CI chondrite-like isotopic composition (Schiller et al., 2018). We have shown that this scenario is inconsistent with (*i*) the lack of a clear temporal trend of the isotopic composition of the NC reservoir towards more CC-rich compositions, (*ii*) the isotopic composition of Earth in the framework of the NC-CC dichotomy of the Solar System, (*iii*) the close isotopic similarity between Earth, aubrites, and enstatite chondrites, and (*iv*) the lack of a CC contribution to Mars together with the overall low CC fraction in Earth. Together, these observations indicate that CC dust from the outer disk never penetrated in significant amount into the terrestrial planet-forming inner disk. Thus, if planets accreted from pebbles, they did so from a reservoir of local dust, presumably trapped in a pressure bump of the disk.

However, even this scenario faces difficulties. The gradual depletion of the BSE in volatile elements of decreasing condensation temperature is in general difficult to reconcile with pebble accretion, which should lead to a quasi-stepwise depletion pattern (i.e. full retention/depletion of elements with condensation temperature above/below a certain value). Moreover, the rapid formation of the proto-Earth within the lifetime of the disk, required in the pebble accretion model, would have inevitably led to a much more radiogenic $^{182}$W composition of the BSE than observed. A single late giant impact, as is commonly invoked for the origin of the Moon, is insufficient for removing this radiogenic $^{182}$W to Earth's core because of the small degree of equilibration between the impactor core and Earth's mantle expected for giant impacts. Moreover, substantial equilibration during the Moon-forming event is not only unsupported by models and experiments but within the pebble accretion model is also inconsistent with the Earth's Ni isotope composition (Onyett et al., 2023).

Geochemical considerations based on the He/Ne ratio argue that the Earth suffered not one but at least two giant impacts during its growth history. Mercury and Mars probably also experienced giant impacts. Moreover, the lunar and martian crater records support the existence of a substantial population of planetesimals, continuously accreting onto the terrestrial planets. All these lines of evidence indicate that the formation of terrestrial planets by mutual collisions among planetary embryos and planetesimals is the correct model. This scenario can easily account for the isotopic, chemical and chronological constraints enumerated above.

**Appendix**

To draw Fig. 6 we assume a target and impactor with equal core mass fractions $g$ and masses $m_t$ and $m_i$, respectively. The impactor/target mass ratio is $g$. The target core mass $m_t^c$ is then

$$m_t^c = \frac{\gamma}{1-\gamma} m_t^m$$

Here and throughout superscripts distinguish mantle and core and subscripts denote the body of interest.

During a collision we assume that all the impactor mantle and a fraction $k$ of the impactor core equilibrates with the target mantle. Given the initial concentrations of some tracer (which could be an isotope anomaly) in the target and impactor, we can use mass balance to write

$$C_t^m m_t^m + C_i^m m_i^m + kC_i^c m_i^c = C_f^m(m_t^m + m_i^m) + kC_f^c m_i^c$$

where here the subscripts *t*, *i* and *f* denote the target, impactor and combined body immediately after the impact. The left-hand side is pre-impact and the right-hand side is post-impact.

Dividing through by $m_t^m$ we obtain

$$C_t^m + gC_i^m + kC_i^c \frac{\gamma}{1-\gamma} g = C_f^m(1+g) + kC_f^c \frac{\gamma}{1-\gamma} g$$

Now if our impactor has a bulk chondritic composition $C_c$ then

$$C_i^m + C_i^c \frac{\gamma}{1-\gamma} = C_c \frac{1}{1-\gamma}$$

So we can write

$$C_t^m + gC_i^m + kg\left(\frac{C_c}{1-\gamma} - C_i^m\right) = C_f^m(1+g) + kC_f^c \frac{\gamma}{1-\gamma} g$$

During the impact there will be partitioning between metal and silicate, with $C_f^c = DC_f^m$ with *D* the partition coefficient. This gives us

$$C_t^m + gC_i^m(1-k) + kg\frac{C_c}{1-\gamma} = C_f^m\left(1 + g + \frac{Dk\gamma g}{1-\gamma}\right)$$

We then explicitly assume that the concentrations represent isotopic anomalies (e.g. ε$^{182}$W). Defining this quantity relative to chondritic, we then have $C_c$=0. Here we are assuming that the impact is sufficiently late that no post-impact radiogenic ingrowth occurs.

If we further assume that the pre-impact target and impactor tungsten anomalies are the same ($C_t^m = C_i^m$) then we can write

$$\frac{C_f^m}{C_t^m} = \frac{(1+g(1-k))(1-\gamma)}{(1+g)(1-\gamma) + Dk\gamma g}$$

The left-hand side is the reduction factor of the initial tungsten anomaly due to the impact. Increasing *D*, the impactor size *g* or the equilibration factor *k* all reduce the final tungsten anomaly. In the limiting case *k*=0 (no equilibration), we find no reduction, as required.

**Acknowledgments**

A.M. and T. K. acknowledge the support from the ERC (project number 101019380-HolyEarth). F.N. acknowledges the support of NASA-EW (80NSSC21K0388). We are thankful to both reviewers for their constructive comments that improved this manuscript.